\documentclass[sigconf,nonacm]{acmart}

\AtBeginDocument{%
  \providecommand\BibTeX{{%
    \normalfont B\kern-0.5em{\scshape i\kern-0.25em b}\kern-0.8em\TeX}}}

\usepackage{algorithmic}
\usepackage{algorithm}
\usepackage[tight,footnotesize]{subfigure}
\usepackage{tabularx}
\usepackage{enumitem}

\begin{document}


\title{UNSEEN: A Cross-Stack LLM Unlearning Defense\\ against AR-LLM Social Engineering Attacks}





%

\author{
Tianlong Yu$^{1}$,
Yang Yang$^{1}$,
Xiao Luo$^{1}$,
Lihong Liu$^{1}$,
Fudu Xing$^{2}$,\\
Zui Tao$^{3}$,
Kailong Wang$^{3}$,
Gaoyang Liu$^{3}$,
Ting Bi$^{3}$
}

\affiliation{
  $^{1}$Hubei University, Wuhan, China \\
  $^{2}$University of Southern California, Los Angeles, USA \\
  $^{3}$Huazhong University of Science and Technology, Wuhan, China\\
tommyyu21@163.com, 
yangyang@hubu.edu.cn,
(xiaoluo, 202421121013103)@stu.hubu.edu.cn,
fuduxing@usc.edu,\\
(tzzz1, wangkl, 2025010019)@hust.edu.cn,
ting.bi@ieee.org
\country{}
}





\begin{abstract}

Emerging AR-LLM-based Social Engineering attack (e.g., SEAR) is at the edge of posing great threats to real-world social life. 
In such AR-LLM-SE attack, the attacker can leverage AR (Augmented Reality) glass to capture the image and vocal information of the target, using the LLM to identify the target and generate the social profile, using the LLM agents to apply social engineering strategies for conversation suggestion to win the target's trust and perform phishing afterwards.
Current defensive approaches, such as role-based access control or data flow tracking, are not directly applicable to the convergent AR-LLM ecosystem (considering embedded AR device and opaque LLM inference), leaving an emerging and potent social engineering threat 
that existing privacy paradigms are ill-equipped to address.
This necessitates a shift beyond solely human-centric measures like legislation and user education toward enforceable vendor policies and platform-level restrictions. 
Realizing this vision, however, faces significant technical challenges: securing resource-constrained AR-embedded devices, implementing fine-grained access control within opaque LLM inferences, and governing adaptive interactive agents. To address these challenges, we present UNSEEN, a coordinated cross-stack defense that combines an AR ACL (Access Control Layer) for identity-gated sensing, F-RMU-based LLM unlearning for sensitive profile suppression, and runtime agent guardrails for adaptive interaction control.
We evaluate UNSEEN in an IRB-approved user study with 60 participants and a dataset of 360 annotated conversations across realistic social scenarios (e.g., coffee shops and networking events), using SEAR as the attack baseline.
The experiment shows that UNSEEN reduces the Social-Engineering attack effectiveness by at least 61\% for phishing photo links, social apps, SMS, and phone calls, providing practical, cross-stack defense against AR-LLM social-engineering attacks.

\end{abstract}

\ccsdesc[500]{Human-centered computing~Human computer interaction (HCI)}

\keywords{Augmented Reality, Multimodal LLMs, Social Engineering Attacks.}


\maketitle



\section{Introduction}
\label{sec:introduction}

The convergence of Augmented Reality (AR) and Large Language Models (LLMs) has ushered in a new era of immersive computing~\cite{arattackmeta,yang2025socialmind, li2024satori, tsai2024gazenoter}, yet it simultaneously enables a novel and potent form of digital threat: the AR-LLM-based Social Engineering (AR-LLM-SE) attack~\cite{sear,seardataset}. In this emerging paradigm, an attacker can leverage AR glasses to surreptitiously capture a target's visual and vocal data, employ an LLM to analyze this data for identity and social profiling, and subsequently deploy LLM-driven interactive agents to craft and suggest highly personalized, trust-building conversation strategies, ultimately facilitating sophisticated phishing or manipulation. This attack vector stands at the edge of posing grave threats to real-world social interactions and personal security, moving cyber-physical threats directly into the fabric of daily life~\cite{iqbal2023adopting}.

Current defensive paradigms are fundamentally ill-suited to this convergent threat. Traditional privacy mechanisms, including role-based access control~\cite{smartauth,contexiot,he2018contextpolicy} and data-flow tracking~\cite{flowfence}, are largely designed for single-layer systems, whereas AR-LLM social engineering operates as a cross-stage pipeline. In this setting, resource-constrained always-on AR devices and opaque, data-hungry model inference create a practical security gap~\cite{sear} that cannot be closed by human-centric countermeasures alone (e.g., legislation or user education).
The key issue is architectural mismatch: device-side controls may limit sensor invocation but cannot prevent downstream profile inference once data is captured; model-side controls are difficult to enforce at identity granularity; and output-time filtering is brittle under adaptive multi-turn prompting. Effective mitigation must jointly constrain sensing, inference, and interaction rather than treating them as independent surfaces.

To address this challenge, we propose UNSEEN, a coordinated cross-stack defense with three complementary layers: (1) \textbf{LLM Unlearning for Sensitive Profile Suppression}, using F-RMU (Fisher-Weighted Sparse Representation Misalignment) to reduce retention and inference of protected social-profile attributes; (2) \textbf{AR ACL for Identity-Gated Sensing}, a reference-monitor-based layer that enforces lightweight, fine-grained mediation on AR devices before data is forwarded downstream; and (3) \textbf{Agent Guardrails for Adaptive Interaction Control}, a runtime, state-aware control layer that constrains adaptive multi-turn responses to block social-engineering policy violations. Detailed workflow in Section~\ref{sec:overview}.

Our main contributions are as follows:
\textbf{Contribution 1: First end-to-end defense against AR-LLM-based Social Engineering attacks}. We present UNSEEN - the first end-to-end defense against AR-LLM social-engineering attacks that protects cross-layer surface spanning AR sensing, LLM inference, and Agent interaction;
\textbf{Contribution 2: A novel cross-stack defense architecture}: We design UNSEEN as an integrated stack that combines AR ACL for identity-gated sensing, F-RMU-based LLM unlearning for sensitive profile suppression, and runtime agent guardrails for adaptive interaction control.
\textbf{Contribution 3: Prototype implementation and IRB dataset}: We implement the full pipeline and constructed an IRB dataset with 60 participants and 360 annotated conversations, and show, through comparative and ablation evaluations, that the proposed stack substantially reduces social-engineering effectiveness while maintaining practical system usability.




\noindent\textbf{IRB Permission:} This study was approved by the IRB. All human-related data were collected under rigorous ethical guidelines, and anonymized prior to analysis, and handled in strict accordance with data protection protocols. No personally identifying information is disclosed in this study. The study adhered to all applicable legal and ethical standards for research involving human subjects. 



\noindent\textbf{UNSEEN Code and Dataset available at:} \url{https://gitlab.com/yqts-group/unseenxxx.git}.
\section{Related Works and Motivation}
\label{sec:motivation}

\begin{table}[t]
    \centering
    \caption{Comparison across three defense dimensions.}
    \label{tab:cross_stack_comparison}
    \begin{tabular}{@{}lccc@{}}
        \toprule
        Approach & Vendor & Coverage & Resilience \\
        \midrule
        AR ACL & AR Glass & Device sensing & Medium \\
        LLM Unlearn & LLM Model & Model inference & Medium \\
        Agent Guardrails & Agent & Interaction layer & Low \\
        UNSEEN & Any & All layers & High \\
        \bottomrule
    \end{tabular}
\end{table}


\subsection{AR-LLM-based Social Engineering Attack}

Recent works show how social engineering can be systematically strengthened by combining AR sensing, LLM reasoning, and interactive Agent. Classical social-engineering studies~\cite{ho2019detecting, bilge2009all, roy2024chatbots, timkounderstanding} mostly focus on human factors and persuasion strategies~\cite{burda2024cognition, vadrevu2019you, yang2023trident, ulqinaku2021real}, while recent LLM-agent works~\cite{falade2023decoding} demonstrate that LLMs can automate context-aware phishing scripts at scale. AR-assisted intelligence systems~\cite{yang2025socialmind,jansen2020social,fuste2017artextiles,hirskyj2020social} show that AR sensors can continuously capture visual and auditory cues from physical environments, dramatically increasing the fidelity of personal context extraction. SEAR \cite{sear,seardataset} connects these threads into an end-to-end attack pipeline, using AR sensing for target capture, multimodal LLM inference for identity/profile construction, and agentic dialogue planning for adaptive trust manipulation, moving social engineering from static message crafting to real-time closed-loop interaction. This emerging attack class motivates defenses that operate across all three stages rather than at a single point defense.

\subsection{AR ACL}
Prior AR/wearable defenses emphasize permission gating, local access control, and on-device recognition to reduce unauthorized camera/microphone use~\cite{roesner2021security,chen2018case,lehman2022hidden}. Yet mobile/AR permission models remain coarse-grained (app-level allow/deny) and typically do not enforce identity-aware capture constraints during live interaction~\cite{felt2011android}. At the same time, edge-efficient vision and open-set recognition indicate that lightweight identity screening is feasible on resource-constrained devices for real-time filtering before raw multimodal signals are forwarded downstream~\cite{chen2018mobilefacenetsefficientcnnsaccurate,geng2021recent}. For AR-LLM social-engineering threats, two gaps remain: permission-centric controls do not address the attacker’s semantic objective (harvesting personally identifying context in open physical environments), and partial sensing constraints are insufficient when downstream profile inference and dialogue generation remain unconstrained.


\subsection{LLM Unlearn}

In Multimodal LLMs, balancing computational efficiency with unlearning efficacy is critical. Among existing schemes, methods based on Low-Rank Adaptation (LoRA) \cite{hu2022lora} have become dominant due to their ability to adapt models with minimal overhead. Several works have adapted LoRA for unlearning tasks; for instance, Forget-Me-Not \cite{zhang2024forget} utilizes LoRA to efficiently update attention weights in text-to-image models to remove specific concepts. 
Concurrently, another class of approaches focuses on modifying model components through sparse updates. Meng et al. \cite{meng2022locating} proposed MEMIT, which locates and edits factual associations in the Feed-Forward Network (FFN) layers of LLMs, while Task Vectors \cite{ilharco2022editing} demonstrated that arithmetic operations on model weights can effectively remove task-specific capabilities. More recently, Yoon et al. \cite{yoon2024few} advanced this paradigm in the context of security, utilizing sparsity-constrained optimization to achieve effective few-shot unlearning.

Despite these advancements, applying these techniques to Multimodal Large Language Models (MLLMs) reveals three coupled limitations: first, most existing approaches \cite{meng2022locating, xing2025continuous, wu2023depn} rely on homogeneous importance metrics (e.g., gradient magnitude or activation frequency), which are effective for discrete text tokens but fail to capture the geometric structure of the vision encoder, where sensitive concepts are distributed over a continuous feature manifold and encoded by local curvature; second, current sparse unlearning methods often use \textit{destructive} edits such as zeroing neurons or directly modifying pre-trained parameters \cite{wu2023depn}, which, as suggested by weight-manipulation analyses in Ilharco et al. \cite{ilharco2022editing}, can disrupt delicate cross-modal feature correlations and trigger a ``lobotomy-style'' degradation of alignment; third, standard unlearning objectives can be numerically unstable in high-dimensional multimodal optimization, so without robust constraints they may diverge, motivating robust losses with sparsity-constrained updates\cite{yoon2024few}.

\subsection{Agent Guardrails}

Related work increasingly shows that the conversational layer is a primary abuse surface for LLM-powered agents~\cite{wang2019exploring, yao2023react, afane2024next, chen2024pandora}. Recent studies~\cite{roy2024chatbots, afane2024next, chen2024pandora} demonstrate that LLM agents can be steered to generate persuasive phishing or jailbreak-enabled social-engineering content with high contextual specificity, and SEAR~\cite{sear} further indicates that this risk becomes stronger when dialogue generation is coupled with AR-based personal-context capture. In response, guardrail-oriented defenses in practice commonly adopt runtime policy checks, refusal templates, and constrained response rewriting to block unsafe disclosures before user delivery. However, most existing guardrails are largely text-rule-centric and single-turn, making them brittle against adaptive multi-turn attacks (e.g., aliasing, prompt reframing, or profile-based inference). These limitations motivate our Agent ACL design, which treats guardrails as a dynamic, state-aware control layer. 

\begin{figure*}[t]
\centering
\includegraphics[width=0.8\textwidth]{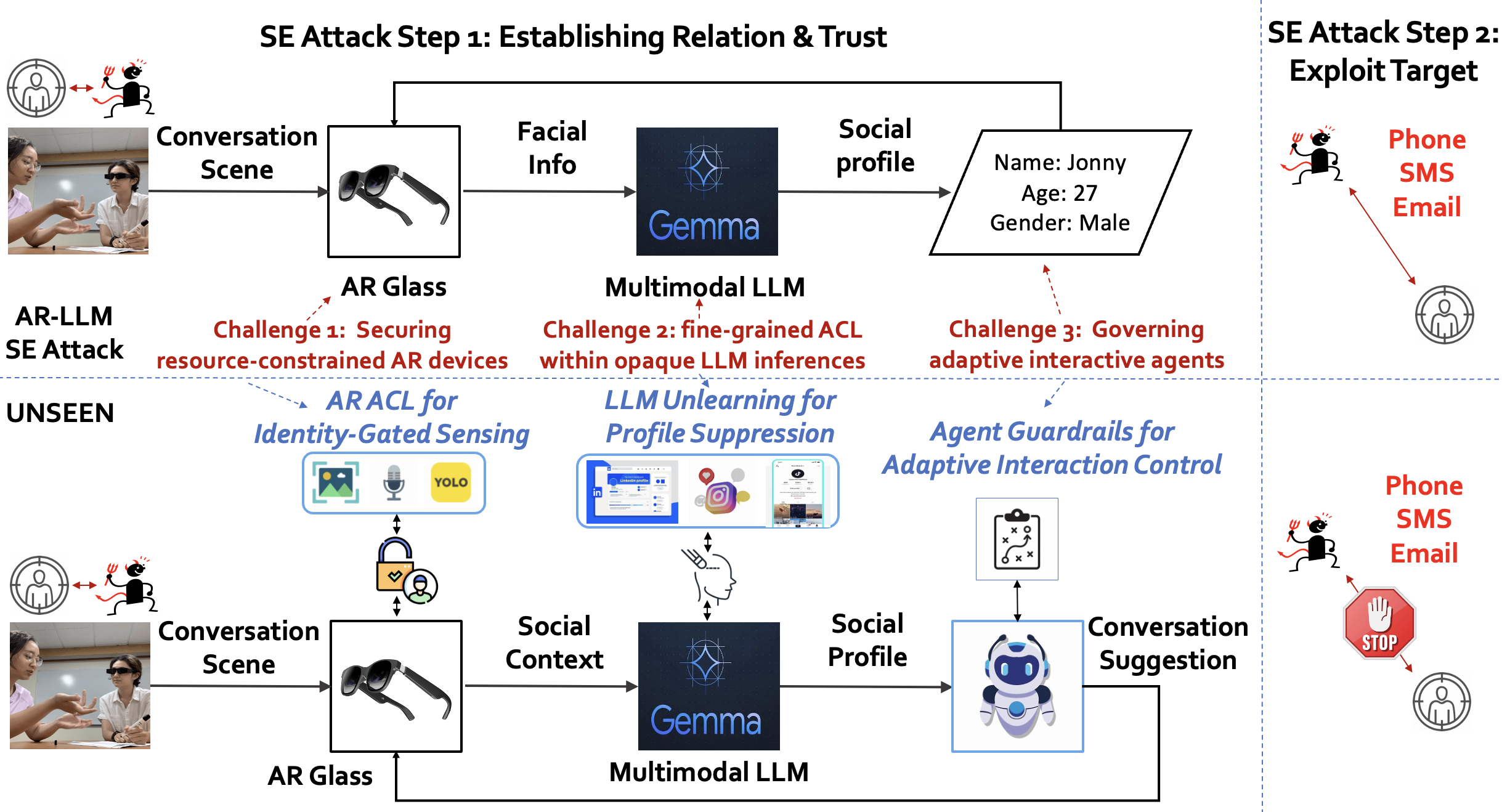}
\caption{AR-LLM Social Engineering attacks (above) and how UNSEEN can prevent it (below).}
\vspace{-10pt}
\label{fig:overview_unseen_arch}
\end{figure*}

\subsection{Need for Cross-Stack Defense}

AR-LLM social engineering is fundamentally a \emph{pipeline attack}: attackers first capture context at the AR sensing layer, then infer sensitive profiles at the model layer, and finally operationalize manipulation through adaptive agent interaction. This attack structure creates a direct mismatch with single-layer defenses, because protection at one stage still leaves two downstream paths available for abuse.
Table~\ref{tab:cross_stack_comparison} shows why UNSEEN is intentionally designed as a cross-stack defense across three key dimensions. \textbf{Vendor requirement:} each single-layer method depends on a specific vendor (AR device, model provider, or agent platform), which limits deployability in heterogeneous ecosystems; UNSEEN is deployable when any one participating vendor can enforce part of the pipeline. \textbf{Coverage breadth:} AR ACL, LLM unlearning, and agent guardrails each protect only one stage, while UNSEEN covers device sensing, model inference, and interaction jointly. \textbf{Bypass resilience:} single-layer controls are easier to route around (e.g., rooted-device capture, prompt adaptation, or fallback pathways), whereas coordinated controls force attackers to evade multiple coupled checkpoints.
Therefore, the motivation for UNSEEN is not additive feature stacking but risk-structure alignment: a cross-layer attack requires cross-layer containment. By combining AR ACL, LLM unlearning, and agent guardrails into one coordinated path, UNSEEN reduces single-point failure and provides stronger defense than any isolated mechanism.

\section{Overview}
\label{sec:overview}


In this section, we provide the overview of UNSEEN, including the threat model, attack stages, defense gap, and cross-stack defense workflow and modules.

\textbf{\textit{Threat model:}} We assume the following adversarial capabilities and deployment conditions:
\begin{itemize}
    \item Adversaries can use commodity AR hardware (camera and microphone) to continuously capture multimodal signals, including facial, vocal, and contextual cues.
    \item Adversaries can aggregate auxiliary social information (e.g., publicly available online profiles) to build target-specific context for personalization.
    \item Adversaries can use LLM-based reasoning and agentic dialogue generation to adapt conversational strategies.
    \item Targets can be impacted by social-engineering factors such as authority bias, reciprocity, and cognitive overload during live interaction.
    \item Current commercial AR platforms do not consistently disable identity-aware capture (e.g., face recording).
\end{itemize}

\textbf{\textit{AR-LLM Social-Engineering attack stages:}} As shown in Figure~\ref{fig:overview_unseen_arch}, attack proceeds as a closed loop: (i) AR sensing captures multimodal target signals, (ii) LLM-based inference extracts identity and social-profile attributes, and (iii) an interactive agent generates adaptive, trust-building dialogue for phishing or manipulation. This coupling of physical-world sensing with model-driven persuasion elevates social engineering from static message crafting to real-time cyber-physical manipulation.

\textbf{\textit{Why current defenses fail:}} Existing defenses are mostly single-layer and therefore mismatched to this cross-layer pipeline. Device-side permissions are often coarse-grained and app-centric, model-side internals are opaque and difficult to constrain semantically, and output filtering alone is brittle under adaptive multi-turn prompting. As a result, sensitive information can still be captured upstream, inferred midstream, and leaked downstream.

Realizing an effective defense therefore requires solving three coupled technical challenges:

\begin{itemize}
    \item \textbf{Challenge 1: Fine-grained ACL within opaque LLM inference.}
    Enforcing access control within opaque generative inference is difficult, sensitive social attributes can be implicitly inferred from benign inputs.

    \item \textbf{Challenge 2: Securing resource-constrained AR devices.}
    AR endpoints must provide robust capture-time sensor governance under strict latency and compute constraints.

    \item \textbf{Challenge 3: Governing adaptive interactive agents.}
    Runtime control of adaptive multi-turn responses is needed to prevent policy evasion and malicious SE strategies.
\end{itemize}

To address these challenges, as shown in Figure~\ref{fig:overview_unseen_arch}, UNSEEN adopts a coordinated three-layer defense stack: 1) \textbf{LLM Unlearning for Sensitive Profile Suppression}: An F-RMU-based unlearning mechanism that reduces retention and inference of protected social-profile attributes in the model;
2) \textbf{AR ACL for Identity-Gated Sensing}:
A reference-monitor-based AR control layer that performs lightweight, fine-grained mediation of sensor access before signals are released downstream;
3) \textbf{Agent Guardrails for Adaptive Interaction Control}:
A runtime, state-aware guardrail layer that monitors and constrains adaptive multi-turn outputs to block malicious social-engineering behaviors.
Together, these three layers provide complementary protection across sensing, inference, and interaction, reducing both upstream exposure and downstream leakage risk.
We present the LLM-layer defense (LLM Unlearning) in Section~\ref{sec:llmunlearn}, and the non-LLM control layers (AR ACL and Agent Guardrails) in Section~\ref{sec:arandagent}.

\section{LLM Unlearn for Profile Suppression}
\label{sec:llmunlearn}

This section presents the LLM-layer defense in UNSEEN, which removes socially sensitive identity knowledge from multimodal models while preserving normal assistant utility. We focus on a targeted unlearning objective: erase attacker-useful persona cues that enable AR-LLM social engineering, without degrading benign recognition and generation performance. To this end, we introduce a structured and stable unlearning pipeline that localizes sensitive parameters and applies minimal, controlled updates.

\begin{figure}[t]
    \centering
    \includegraphics[width=\linewidth]{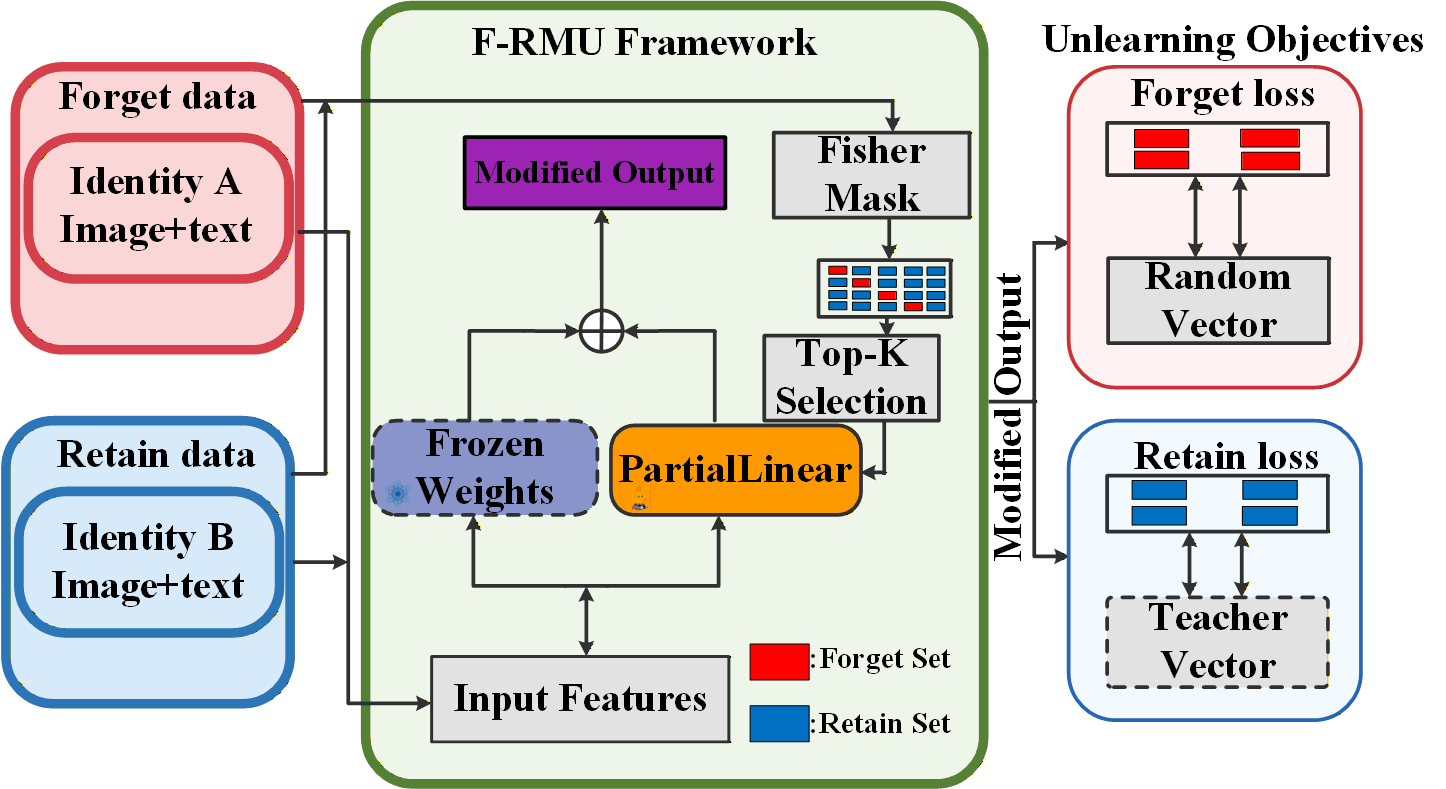} 
    \caption{F-RMU framework.}
    \vspace{-10pt}
    \label{fig:unlearnframework}
\end{figure}

\begin{algorithm}[t]
\caption{Fisher-Weighted Sparse Representation Misalignment}
\label{alg:frmu}
\begin{algorithmic}[1]
\REQUIRE Pre-trained LMM $\theta$, Forget Set $D_f$, Retain Set $D_r$
\REQUIRE Integration steps $m$, Top-$K$ sparsity ratio $k$, Hyperparameters $\beta, \gamma$
\ENSURE Safety-Aligned Unlearned Model $\theta^*$

\STATE \textbf{Phase 1: Heterogeneous Geometric Localization}
\STATE Initialize accumulators $I_{fisher} \leftarrow 0, I_{grad} \leftarrow 0$
\FOR{step $s = 1$ to $m$}
    \STATE $\alpha \leftarrow s/m$, $\theta' \leftarrow \alpha \theta$ \COMMENT{Interpolation on Manifold}
    \STATE Compute Gradients $G \leftarrow \nabla_{\theta'} \mathcal{L}(D_f; \theta')$
    \STATE $I_{fisher} \leftarrow I_{fisher} + (G \odot G)$ \COMMENT{Accumulate Diagonal Fisher}
    \STATE $I_{grad} \leftarrow I_{grad} + |G|$ \COMMENT{Accumulate Integrated Gradients}
\ENDFOR
\STATE Calculate Scores: $S \leftarrow \frac{1}{m}(I_{fisher} \cdot \theta^2 \oplus I_{grad} \cdot |\theta|)$
\STATE Identify Attack Surface: $\mathcal{S} \leftarrow \text{TopK}(S, k)$

\STATE \textbf{Phase 2: Structured Sparse Setup}
\STATE Freeze backbone $\theta_{\text{frozen}} \leftarrow \theta$
\STATE Initialize sparse residuals $\Delta W_{\mathcal{S}} \leftarrow 0$ \COMMENT{Zero-Init Stability}

\STATE \textbf{Phase 3: Robust RMU Optimization}
\STATE Define Teacher $\mathcal{M}_{\text{tea}} \leftarrow \text{Copy}(\theta)$
\WHILE{not converged}
    \FOR{batch $(x_f, x_r)$ in $(D_f, D_r)$}
        \STATE Extract features $H_{\text{stu}}(x)$, $H_{\text{tea}}(x)$
        \STATE \textit{Obfuscation Step (Forget)}:
        \STATE $v_{rand} \sim \mathcal{N}(0, I)$, $\gamma \leftarrow \| H_{\text{tea}}(x_f) \|_2$ 
        \STATE $\mathcal{L}_{\text{forget}} \leftarrow \text{Huber}(H_{\text{stu}}(x_f), \gamma \cdot v_{rand})$
        \STATE \textit{Preservation Step (Retain)}:
        \STATE $\mathcal{L}_{\text{retain}} \leftarrow \text{Huber}(H_{\text{stu}}(x_r), H_{\text{tea}}(x_r))$
        \STATE Update $\Delta W_{\mathcal{S}} \leftarrow \Delta W_{\mathcal{S}} - \eta \nabla (\mathcal{L}_{\text{forget}} + \beta \mathcal{L}_{\text{retain}})$
    \ENDFOR
\ENDWHILE

\STATE \textbf{Phase 4: Lossless Merge}
\RETURN $\theta^* \leftarrow \theta_{\text{frozen}} \oplus \Delta W_{\mathcal{S}}$
\end{algorithmic}
\end{algorithm}

\subsection{F-RMU Framework}
As show in Algorithm~\ref{alg:frmu}, we present the \textbf{Fisher-Weighted Sparse Representation Misalignment (F-RMU)}. The goal is targeted unlearning in Large Multimodal Models (LMMs): remove attacker-useful identity concepts while preserving general utility and safety alignment. As shown in Figure~\ref{fig:unlearnframework}, F-RMU follows two coupled stages: (1) \textit{Heterogeneous Geometric Sensitivity Analysis} to localize high-risk parameters, and (2) \textit{Structured Sparse Adaptation} to apply minimal, localized updates for concept removal.
Figure~\ref{fig:unlearnframework} illustrates the full pipeline. On the left, the model receives forget samples (Identity A) and retain samples (Identity B). In the middle, backbone parameters are frozen (blue), and a parallel \textit{PartialLinear} residual branch (orange) is inserted; a Fisher-guided Top-$K$ mask activates only the most sensitive neurons, reducing collateral parameter drift. On the right, F-RMU outputs from frozen and residual paths are merged into a modified representation and optimized with dual objectives: (1) \textit{Forget Loss}, which maps forget representations toward random directions for obfuscation, and (2) \textit{Retain Loss}, which distills from the teacher model to preserve non-target capabilities.

\subsection{Heterogeneous Geometric Sensitivity}
To minimize collateral damage and identify the precise attack surface where sensitive concepts are encoded, we argue that vision and text parameters reside on different optimization manifolds.

\noindent\textbf{\textit{Vision Tower: Integrated Fisher Information (IFI).}}
For the vision encoder, we model parameter importance using Riemannian geometry. 
We derive a path-integrated metric based on the Fisher Information Matrix (FIM).

To quantify the sensitivity of parameter $\theta \in \mathbb{R}^d$ with respect to the forget set $D_f$, we analyze the change in loss $\Delta \mathcal{L}$ under a perturbation $\delta$. Using a second-order Taylor expansion around $\theta$:
\begin{equation}
\Delta \mathcal{L}(\theta + \delta) \approx \mathcal{L}(\theta) + \nabla \mathcal{L}(\theta)^T \delta + \frac{1}{2} \delta^T H(\theta) \delta
\end{equation}
Assuming the model is near convergence ($\nabla \mathcal{L} \approx 0$) and approximating the Hessian $H(\theta)$ with the Fisher Information Matrix $F(\theta)$, the loss change is dominated by the curvature term:
\begin{equation}
\Delta \mathcal{L} \approx \frac{1}{2} \delta^T F(\theta) \delta
\end{equation}
The importance score $S_i$ for removing the $i$-th parameter ($\delta_i = -\theta_i$) is thus $S_i \approx \frac{1}{2} F_{ii}(\theta) \theta_i^2$. To mitigate local fluctuations in curvature, we extend this point estimate to a path integral along the trajectory $\gamma(\alpha) = \alpha \theta^*$ from $\alpha=0$ to $1$:
\begin{equation}
\mathcal{I}_{Riem}(\theta_i) = \int_{0}^{1} F_{ii}(\gamma(\alpha)) \cdot (\gamma_i(\alpha))^2 \, d\alpha
\end{equation}
By substituting the empirical definition of diagonal Fisher as $F_{ii} \approx \mathbb{E}[(\partial \mathcal{L}/\partial \theta_i)^2]$, we derive \textbf{Integrated Fisher Information (IFI)}:
\begin{equation}
\label{eq:ifi_derivation}
\mathcal{I}_{IFI}^{(i)} = \int_{0}^{1} \mathbb{E}_{(x, y) \sim D_f} \left[ \left( \frac{\partial \mathcal{L}(f(\alpha \theta^*))}{\partial (\alpha \theta_i)} \right)^2 \right] \odot (\theta_i)^2 \, d\alpha
\end{equation}

We approximate this integral using a Riemann sum with $m$ steps. This metric captures the global curvature of the loss landscape, identifying neurons structurally critical for the forget set.
Figure~\ref{fig:heatmap} is a heatmap comparing IFI scores vs. simple Gradient Magnitude, which shows that IFI highlights sparse, specific regions while gradients are noisy/diffuse,
which justifies our partial linear design.

\begin{figure}[t]
    \centering
    \includegraphics[width=\linewidth]{figure2_heatmap.png}
    \caption{\textbf{Visualization of Neuron Importance via Fisher Information.} Top: Random baseline showing uniform noise. Bottom: Our Fisher-weighted metric reveals highly \textbf{structured sparse patterns} (vertical bands).}
    \vspace{-15pt}
    \label{fig:heatmap}
\end{figure}

\noindent\textbf{\textit{Text Projector: Integrated Gradients (IG).}}
For the projection layers, we utilize Euclidean attribution via Integrated Gradients to measure the contribution of neurons to the sparse textual output:
\begin{equation}
S_t^{(i)} = |W_{:i}| \odot \int_{\alpha=0}^{1} \left| \nabla_{\alpha W} \mathcal{L}_{\text{text}}(f(\alpha W); x, y) \right| d\alpha
\end{equation}
Based on these metrics, we generate a binary mask $M^{(l)}$ for each layer by selecting top-$K$ neurons with the highest scores.


\subsection{Column-wise Sparse Residual Adaptation}
Existing Parameter-Efficient Fine-Tuning (PEFT) methods like LoRA assume low-rank updates ($\Delta W = A \times B$). We argue that unlearning is inherently a \textbf{sparse update problem}, requiring surgical modification of local circuits rather than global low-rank adjustments. To this end, we introduce \textbf{PartialLinear}, a structured sparsity mechanism.
We decompose the layer computation into a frozen backbone and a trainable sparse residual. For a selected subset of neuron indices $\mathcal{S}$ (derived from the top-$K$ mask), the forward pass is formulated as:
\begin{equation}
y = x W_{\text{frozen}}^T + \mathcal{P}_{\mathcal{S}}(x) \Delta W_{\mathcal{S}}^T
\end{equation}
where $\mathcal{P}_{\mathcal{S}}(\cdot)$ projects the input $x$ onto the subspace of indices $\mathcal{S}$ (column slicing), and $\Delta W_{\mathcal{S}} \in \mathbb{R}^{d_{out} \times K}$ is a trainable residual matrix initialized to zero.

Instead of relying on empirical comparisons with specific baselines, we highlight the intrinsic theoretical merits of our proposed PartialLinear mechanism in a unified view: it provides \textbf{surgical precision} by updating only the columns associated with critical neurons and thus limiting collateral damage to general knowledge, ensures \textbf{optimization stability} because the zero initialization ($\Delta W_{\mathcal{S}}=0$) makes the starting model mathematically identical to the original ($y_{\text{init}}=y_{\text{orig}}$), achieves \textbf{extreme efficiency} by training only a tiny fraction (e.g., Top-1\%) of parameters, supports \textbf{zero-overhead inference} through lossless post-training merge ($W_{\text{final}}=W_{\text{frozen}}+\Delta W_{\mathcal{S}}$), and strengthens \textbf{security assurance} by confining parameter changes to targeted concept circuits instead of perturbing safety-critical regions.

\begin{figure}[t]
    \centering
    \includegraphics[width=0.95\linewidth]{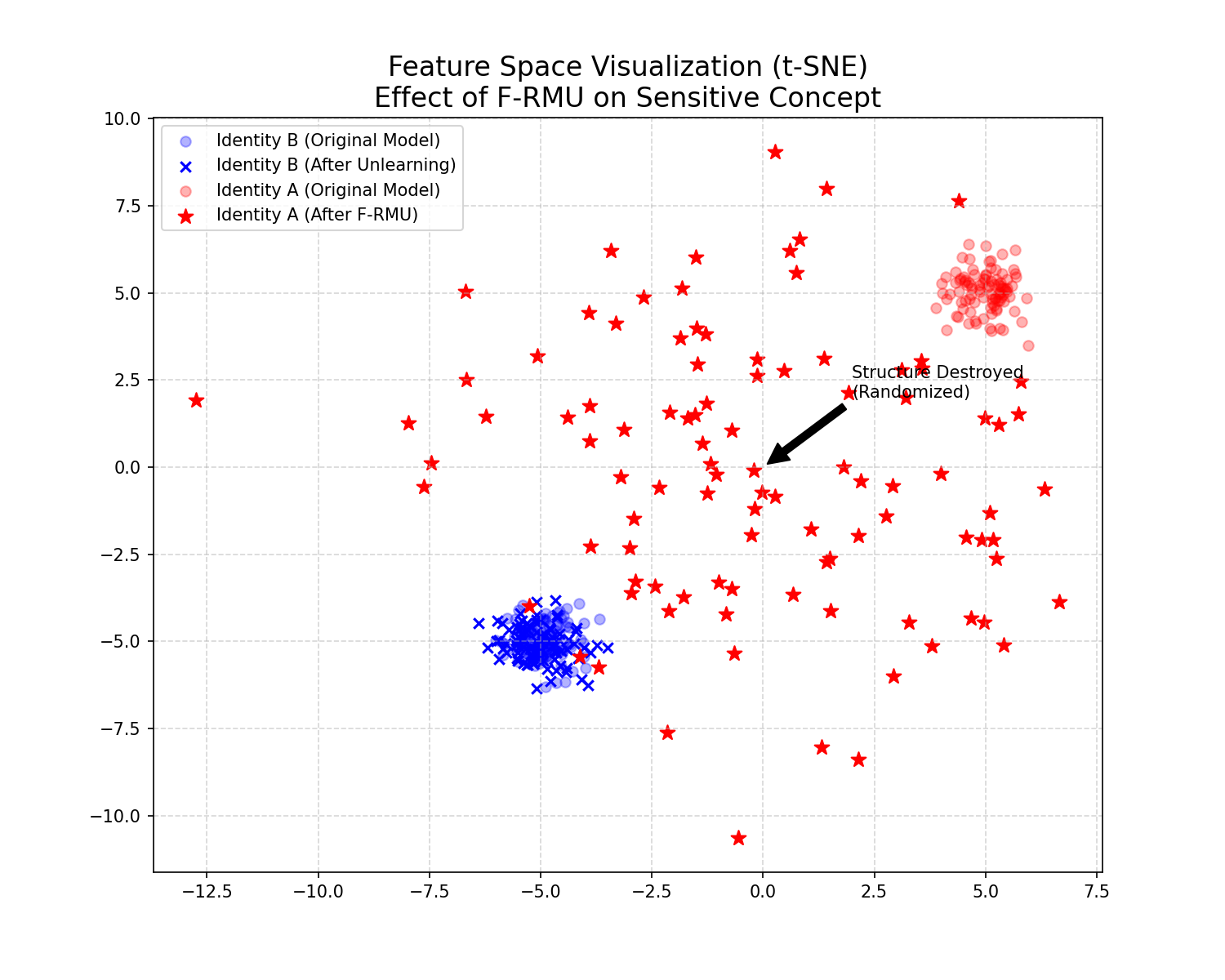} 
    \caption{\textbf{t-SNE Visualization of Feature Manifold Transformation.} The plot shows the embedding space before and after F-RMU. The retained identity (Blue) remains invariant (overlapping clusters), while the target identity (Red) is \textbf{exploded} into a high-entropy distribution, confirming robust privacy obfuscation against inversion attacks.}
    \vspace{-15pt}
    \label{fig:tsne}
\end{figure}

\subsection{Representation Misalignment Optimization}
To erase the target concept, we propose \textbf{Representation Misalignment (RMU)} objective within a teacher-student framework. Let $\mathcal{M}_{\text{student}}$ be trainable model and $\mathcal{M}_{\text{teacher}}$ be the frozen reference.

\noindent\textbf{\textit{Forget Loss: Random Manifold Mapping.}}
For forget samples $x_f \in D_f$, we enforce the student model to map visual features to a random vector $v_{\text{rand}}$ and employ \textbf{Dynamic Norm Matching}:
\begin{equation}
\mathcal{L}_{\text{forget}} = \text{HuberLoss}\left(H_{\text{stu}}(x_f), \gamma \cdot v_{\text{rand}}\right)
\end{equation}
where $\gamma = \| H_{\text{tea}}(x_f) \|_2$ matches the magnitude of the original features. This obfuscates the semantic direction while preserving activation statistics, rendering the target concept indistinguishable from random noise.

\noindent\textbf{\textit{Retain Loss and General Capability.}}
For retain samples $x_r \in D_r$, we use distillation to preserve general capabilities:
\begin{equation}
\mathcal{L}_{\text{retain}} = \text{HuberLoss}\left(H_{\text{stu}}(x_r), H_{\text{tea}}(x_r)\right)
\end{equation}

To empirically validate the efficacy of our method, we visualize the embedding space using t-SNE, as shown in Figure \ref{fig:tsne}. In the original model, the embeddings of the sensitive concept (Identity A, light red circles) form a tightly bound and distinct cluster. However, following the application of F-RMU, these features (red stars) are violently exploded and scattered into a high-entropy distribution. From a security perspective, this thoroughly destroys the semantic structure of the target concept, providing robust privacy obfuscation that makes it computationally infeasible for malicious actors to reconstruct the sensitive identity via model inversion attacks. Conversely, the representations of the retained concept (Identity B) remain remarkably stable, with the embeddings before (blue circles) and after (blue crosses) unlearning perfectly overlapping. This visual evidence confirms that our approach achieves precise, localized erasure and strong privacy guarantees without catastrophic forgetting.
The final objective is $\mathcal{L} = \mathcal{L}_{\text{forget}} + \beta \mathcal{L}_{\text{retain}}$. We explicitly choose \textbf{Huber Loss} over MSE. In high-dimensional feature spaces, outliers are common. Huber Loss transitions to a linear penalty for large errors, providing robustness against gradient explosion. 
\section{AR ACL and Agent Guardrails}
\label{sec:arandagent}

This section presents UNSEEN’s non-LLM control layers as a coordinated defense for the two externally exposed stages of the AR-LLM pipeline: \textbf{input-side AR sensing} and \textbf{output-side Agent interaction}. AR ACL enforces front-end identity-gated sensing on resource-constrained AR devices to reduce unauthorized data capture at the source, while Agent Guardrails enforce release-time policy constraints to prevent protected-identity leakage through adaptive multi-turn dialogue. Jointly, these two layers 
complement model-side unlearning protection by reducing both upstream exposure and downstream disclosure risk.

\subsection{AR ACL for Identity-Gated Sensing}
\label{subsec:aracl_combined}


We model AR ACL as a front-end \textbf{reference monitor} that performs complete mediation before any sensed identity signal is released to downstream LLM modules. Let $x$ denote a detected face crop and $f(\cdot)$ be the embedding encoder on device, producing $z=f(x)$. Let the enrolled whitelist be $\mathcal{W}=\{w_1,\dots,w_n\}$, where each $w_i$ is a protected identity template. The access decision is defined by open-set similarity verification:
\begin{equation}
\hat{i}=\arg\max_i \cos(z,w_i), \qquad
\text{grant iff } \cos(z,w_{\hat{i}})\ge\tau,
\end{equation}
and deny otherwise. This formulation follows least-privilege access control: only sufficiently confident matches to authorized identities can pass the AR boundary.

To balance security and usability, the threshold $\tau$ controls false accept rate (FAR) and false reject rate (FRR), while deployment must satisfy an edge-latency budget, which is a constrained objective:
\begin{equation}
\min_{\tau}\; \lambda\,\mathrm{FAR}(\tau)+(1-\lambda)\,\mathrm{FRR}(\tau)
\quad \text{s.t.}\quad L\le L_{\max},
\end{equation}
where $L$ is end-to-end on-device inference latency. This provides the theoretical basis for our design choice: lightweight open-set verification on AR hardware as the first containment layer.

In our AR ACL realization, faces are first detected and cropped using MediaPipe’s BlazeFace-based detector \cite{lugaresi2019mediapipeframeworkbuildingperception,bazarevsky2019blazefacesubmillisecondneuralface}, followed by a simple 2D alignment step. The aligned face patch is then encoded by the InsightFace buffalosc recognition model, which adopts an MBF/MobileFaceNet backbone from the official model zoo for efficient face-embedding extraction on resource-constrained devices \cite{chen2018mobilefacenetsefficientcnnsaccurate}. 
The AR ACL mechanism is implemented on Rayneo X3 Pro glasses.



\subsection{Agent Guardrails for Interaction Control}
\label{subsec:agentguardrails_combined}

Agent ACL serves as UNSEEN’s release-time enforcement layer. In AR-LLM social-engineering attacks, risk does not end at sensing or model inference; it is operationalized at the interaction endpoint, where the agent produces personalized, persuasive language. Thus, even after AR ACL reduces upstream exposure and LLM unlearning suppresses sensitive internal representations, adaptive prompting and multi-turn context can still recover residual identity cues. Agent ACL addresses this final exposure point by applying policy checks directly to user-visible outputs before release.
This design completes UNSEEN’s cross-stack defense logic: AR ACL constrains what enters the pipeline, LLM unlearning constrains what can be inferred, and Agent ACL constrains what can be disclosed. The release-time layer is therefore indispensable against practical bypass strategies such as aliasing, contextual inference, and conversational steering.

\noindent\textbf{\textit{Context-based dynamic policy enforcement:}}
We formalize agent guardrails as a dynamic release controller over dialogue turns. At turn $t$, the system state is
\begin{equation}
    s_t=(h_t,r_t,c_t),
\end{equation}
where $h_t$ is dialogue history, $r_t \in [0,1]$ is current privacy risk, and $c_t$ is consent/protection context. For candidate output $y_t$, let $\Phi(y_t)$ extract person entities and let $p: \mathcal{E} \rightarrow \{0,1\}$ denote protection labels,
the release policy is
\begin{equation}
\mathcal{R}(y_t,s_t)=
\begin{cases}
\texttt{msg}_{safe}, & \exists e\in\Phi(y_t),\; p(e)=1,\\
\texttt{sanitize}(y_t), & r_t>\tau_t,\\
y_t, & \text{otherwise},
\end{cases}
\end{equation}
with adaptive dynamics
\begin{equation}
    r_{t+1}=f(r_t,y_t,o_t), \qquad \tau_{t+1}=g(\tau_t,\ell_t),
\end{equation}
where $o_t$ is runtime observation (e.g., trigger events) and $\ell_t$ is online error feedback. The corresponding safety invariant is
\begin{equation}
\forall t,\;\forall e\in\Phi(\mathcal{R}(y_t,s_t)),\; p(e)=0,
\end{equation}
which guarantees that protected identities are excluded. 

\noindent\textbf{\textit{From Name Matching to Profile-Similarity ACL:}}
A strict name-matching ACL is bypassable via aliases, misspellings, transliteration variants, or indirect references (e.g., role + affiliation). To improve robustness, we extend policy triggers from name equality to profile similarity.
%
For each protected entity $e$, we maintain a profile representation $z_e$ (multimodal identity features + textual attributes). From current context and candidate output, we construct query profile $q_t$ and compute
\begin{equation}
\text{sim}(q_t,e)=\alpha\,\cos\!\left(q_t^{(v)},z_e^{(v)}\right)+(1-\alpha)\,\cos\!\left(q_t^{(t)},z_e^{(t)}\right),
\end{equation}
where visual and textual channels are jointly weighted. A protected match is triggered when
\begin{equation}
\exists e\in\mathcal{P}:\; \text{sim}(q_t,e)>\delta_t,
\end{equation}
with adaptive threshold $\delta_t$ calibrated from false-positive/false-negative feedback. This makes ACL resilient to lexical obfuscation while remaining compatible with existing guardrail actions. 







\section{Dataset and Methodology}
\label{sec:dataset}

\begin{table*}[t]
\centering
\caption{Social-engineering effectiveness score distribution comparison between SEAR and UNSEEN.}
\vspace{-5pt}
\label{tab:sear_unseen_score_distribution}
\small
\begin{tabular}{@{}lcccccccccccccc@{}}
\toprule
& \multicolumn{7}{c}{SEAR} & \multicolumn{7}{c}{UNSEEN} \\
\cmidrule(lr){2-8} \cmidrule(lr){9-15}
Metric & 5pt(\%) & 4pt(\%) & 3pt(\%) & 2pt(\%) & 1pt(\%) & $\mathbb{E}[\mathrm{Score}]$ & $\sigma(\mathrm{Score})$ & 5pt(\%) & 4pt(\%) & 3pt(\%) & 2pt(\%) & 1pt(\%) & $\mathbb{E}[\mathrm{Score}]$ & $\sigma(\mathrm{Score})$ \\
\midrule
Photo Link & 40.0 & 53.3 & 3.3 & 3.3 & 0.0 & 4.30 & 0.69 & 0.0 & 0.0 & 0.0 & 66.7 & 33.3 & 1.67 & 0.47 \\
Social App & 43.3 & 50.0 & 6.7 & 0.0 & 0.0 & 4.37 & 0.61 & 0.0 & 0.0 & 0.0 & 63.3 & 36.7 & 1.63 & 0.48 \\
SMS & 45.0 & 46.7 & 3.3 & 5.0 & 0.0 & 4.32 & 0.76 & 0.0 & 0.0 & 0.0 & 68.3 & 31.7 & 1.68 & 0.47 \\
Phone Call & 35.0 & 50.0 & 13.3 & 1.7 & 0.0 & 4.18 & 0.72 & 0.0 & 0.0 & 0.0 & 61.7 & 38.3 & 1.62 & 0.49 \\
\midrule
\multicolumn{15}{l}{\footnotesize Score mapping: Very Likely = 5 pt; Likely = 4 pt; Neutral = 3 pt; Unlikely = 2 pt; Very Unlikely = 1 pt.} \\
\bottomrule
\end{tabular}
\vspace{-5pt}
\end{table*}

This section describes how we construct the dataset and evaluation protocol for measuring social-engineering risk and defense effectiveness in AR-LLM interactions, which is also in consistent with SOTA AR-LLM-SE attack setting~\cite{sear, seardataset}.


\subsection{Interaction Scenarios and Data Collection}

\textbf{\textit{Scenario Setup.}} 
We conduct real-world interactions in social settings (e.g., coffee shops and networking events) with 60 participants and 360 total conversations. Participants rotate roles across trials (target vs. attacker) to balance subject-specific bias. Each participant experiences six conditions:
(1) \textbf{Basic conversation} (no technological assistance);
(2) \textbf{SEAR} (AR + multimodal LLM + social agent);
(3) \textbf{UNSEEN (full stack)};
(4) \textbf{UNSEEN (AR ACL only)};
(5) \textbf{UNSEEN (LLM Unlearn only)};
(6) \textbf{UNSEEN (Agent Guardrail only)}.
This design enables both end-to-end comparison (SEAR vs. UNSEEN) and causal attribution of each defense layer through ablation.

\textbf{\textit{Dataset Composition.}} 
The dataset contains four components:
(1) \textbf{AR Multimodal Data}: visual, audio, and environment context captured by AR glasses (e.g., facial/body cues, speech transcripts, time/location/object context);
(2) \textbf{Open Social Data}: publicly available participant-related text/image/video signals used by the attack pipeline;
(3) \textbf{Post-Interaction Survey Data}: structured subjective and behavioral responses;
(4) \textbf{UNSEEN-Protected Outputs}: model outputs and interaction traces after applying full UNSEEN and each single-layer ablation, which provides evidence of how cross-stack defense changes generated social profiles and dialogue behaviors.



\subsection{Survey Questionnaire Design}

\textbf{\textit{Post-Interaction Survey.}} 
All questions use a 5-point Likert scale (1 = Strongly Disagree, 5 = Strongly Agree), unless otherwise noted. The survey is organized into three parts:
\textit{(A) Condition-Level Experience (including UNSEEN ablation):}
Participants rate conversation experience in each condition:
(1) Basic conversation;
(2) SEAR;
(3) UNSEEN (full stack);
(4) UNSEEN (AR ACL only);
(5) UNSEEN (LLM Unlearn only);
(6) UNSEEN (Agent Guardrail only).
This part quantifies the usability-security tradeoff of cross-stack defense and each component;
\textit{(B) Subjective Interaction Quality:}
We collect perception metrics including relevance, appropriateness, naturalness, pacing, sincerity, emotional progression, AR comfort, willingness without AR, future interaction intent, perceived conversational depth, and acceptance of future system use;
\textit{(C) Social-Engineering Susceptibility:}
We measure behavioral intent after each interaction:
(1) clicking shared photo links,
(2) adding the person on social apps,
(3) clicking SMS links,
(4) answering phone calls,
(5) trust before interaction,
(6) trust after interaction.
These items serve as direct proxies for attack success probability and are the primary downstream indicators for UNSEEN effectiveness.

\textbf{\textit{Participant Demographics.}} 
The cohort includes 60 participants aged 23--62 (average age 34), with near-balanced gender distribution (28 male, 32 female) and a variaty of professions. This demographic spread supports evaluating UNSEEN under heterogeneous social interaction styles. More details are avaliable in UNSEEN dataset.

\section{Experiments}\label{sec:simu}


\subsection{Effectiveness against AR-LLM-SE Attacks}

Table~\ref{tab:sear_unseen_score_distribution} shows that UNSEEN substantially reduces SEAR's social-engineering effectiveness across all four attack channels (Photo Link, Social App, SMS, and Phone Call). Under SEAR, high susceptibility responses (``Very Likely'' + ``Likely'') dominate: 93.3\% for Photo Link, 93.3\% for Social App, 91.7\% for SMS, and 85.0\% for Phone Call. In contrast, under UNSEEN, the same actions shift entirely to low-susceptibility responses (``Unlikely'' + ``Very Unlikely''), with 0.0\% in ``Very Likely''/``Likely'' across all four metrics.
This distributional shift is also reflected in the expected score. The mean score drops from 4.30 to 1.67 for Photo Link, from 4.37 to 1.63 for Social App, from 4.32 to 1.68 for SMS, and from 4.18 to 1.62 for Phone Call. These correspond to relative reductions of approximately 61.2\%, 62.7\%, 61.1\%, and 61.2\%, respectively. Overall, UNSEEN consistently pushes participant responses from likely compliance to likely rejection.
These results indicate that UNSEEN effectively mitigates SEAR's practical attack success by reducing both the central tendency and the high-risk response mass in each communication channel. The findings support UNSEEN as an effective defensive mechanism against AR-LLM-based Social-Engineering attacks.

\subsection{UNSEEN Latency Metrics}

Table~\ref{tab:evaluation_latency} reports Min, Max, P90, and Average latency for SEAR and UNSEEN across 60 conversations. We use P90 to characterize tail latency, i.e., the value below which 90\% of observations fall.
For SEAR, AR is lightweight (80.6 ms average; 92.4 ms P90), Social Agent introduces moderate per-stage delay (2.8 s average; 4.0 s P90), and the Multimodal LLM dominates latency (43.3 s average; 52.7 s P90) due to profile generation.
For non-LLM protection, UNSEEN adds explicit protection costs at the AR and agent stages: AR ACL averages 612.1 ms (420.8 ms P90), and Agent Guardrail averages 12.5 s (21.7 s P90). 
However, a counter-intuitive but important result is that UNSEEN's LLM Unlearn module (22.3 s average) is substantially faster than SEAR's Multimodal LLM module (43.3 s average). This comes from architectural differences: UNSEEN uses a more direct model-internal profile generation path, whereas SEAR depends on slower retrieval-heavy profile construction (e.g., RAG-style lookup)~\cite{sear}.
When aggregated across the three modules, SEAR's average pipeline latency is about 46.2 s, while UNSEEN's is about 45.4 s, yielding a net reduction of approximately 1.7\%.
Overall, these results show that stronger security in UNSEEN does not necessarily require higher end-to-end latency.

\begin{table}[t]
\small
\centering
\caption{Latency comparison between SEAR and UNSEEN.}
\vspace{-5pt}
\label{tab:evaluation_latency}
\begin{tabular}{@{}llcccc@{}}
\toprule
Approach & Component & Min & Max & P90 & Average \\
\midrule
SEAR & AR & 64.5ms & 95.1ms & 92.4ms & 80.6ms \\ 
 & Multimodal LLM & 30.2 s & 54.9 s & 52.7 s & 43.3 s \\
 & Social Agent & 1.0 s & 10.6 s & 4.0 s & 2.8 s \\
\midrule
UNSEEN & AR ACL & 312.7ms & 741.3ms & 420.8ms & 612.1ms \\ 
 & LLM Unlearn & 19.0 s & 32.4 s & 24.4 s & 22.3 s \\ 
 & Agent Guardrail & 6.4 s & 22.6 s & 21.7 s & 12.5 s  \\ 
\bottomrule
\end{tabular}
\vspace{-5pt}
\end{table}

\begin{figure}[t]
\centering
\includegraphics[width=0.48\textwidth]{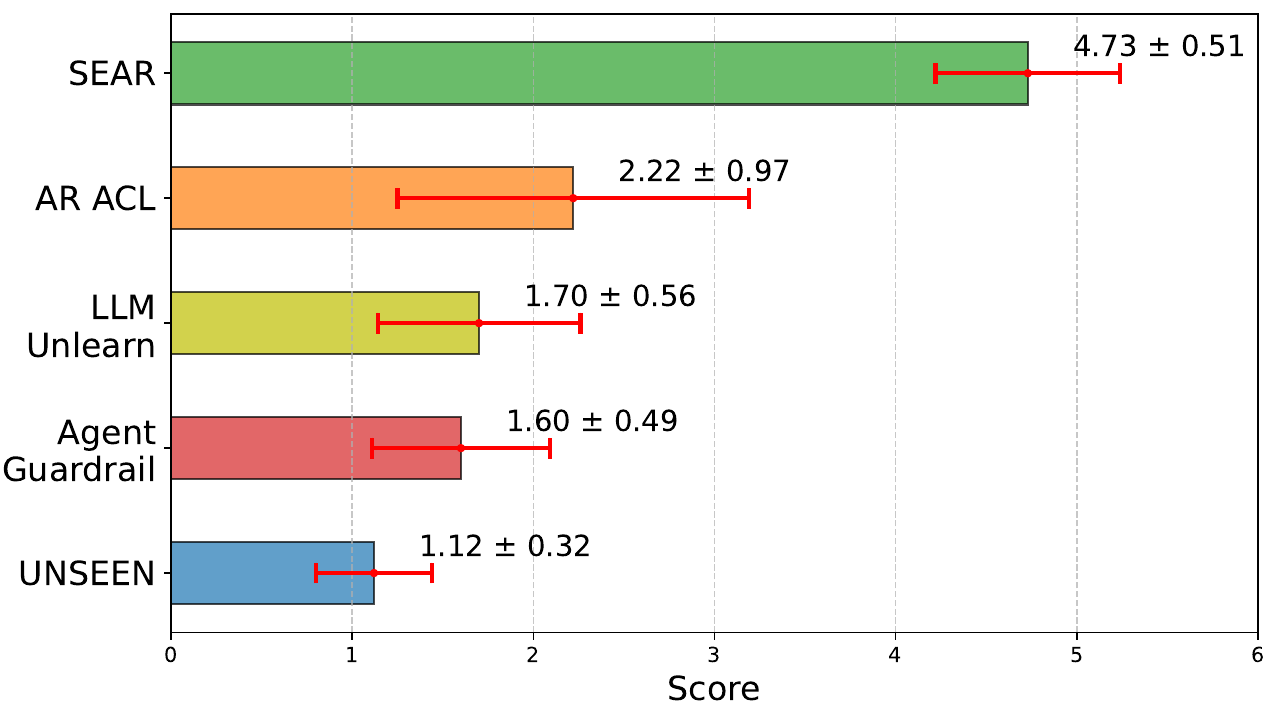}
\caption{UNSEEN ablation study via average social experience score (bar) and standard deviation(errorbar).}
\label{fig:experiment_baseline}
\vspace{-15pt}
\end{figure}

\subsection{UNSEEN Ablastion Study}

Figure~\ref{fig:experiment_baseline} presents the ablation study of UNSEEN by progressively removing key defense modules and comparing average social experience score (bar) and standard deviation(errorbar). The evaluated settings are: UNSEEN (all modules enabled), Agent Guardrail removed, LLM Unlearn removed, AR ACL removed, and SEAR (no UNSEEN defense). Scores are computed from the baseline-comparison questions in Section~\ref{sec:dataset}, where lower scores indicate lower social-engineering success.
The results show a clear monotonic trend: as defensive components are removed, SEAR's effectiveness recovers. UNSEEN achieves the lowest mean score at 1.12 ($\pm$0.32). Removing Agent Guardrail increases the score to 1.60 ($\pm$0.49), removing LLM Unlearn further increases it to 1.70 ($\pm$0.56), and removing AR ACL causes the largest degradation among ablations, reaching 2.22 ($\pm$0.97). Without UNSEEN defenses (SEAR), the mean score rises to 4.73 ($\pm$0.51).
Compared with SEAR, full UNSEEN reduces the average effectiveness score by 76.3\%. The ablation gaps also quantify each module's contribution: +0.48 after removing Agent Guardrail, +0.58 after removing LLM Unlearn, and +1.10 after removing AR ACL (all relative to full UNSEEN). 
Overall, the ablation study confirms that UNSEEN's defense capability is not produced by a single component; it emerges from the coordinated design of access control, profile-level unlearning, and agent-level guardrails.

\begin{figure}[t]
\centering
\includegraphics[width=0.48\textwidth]{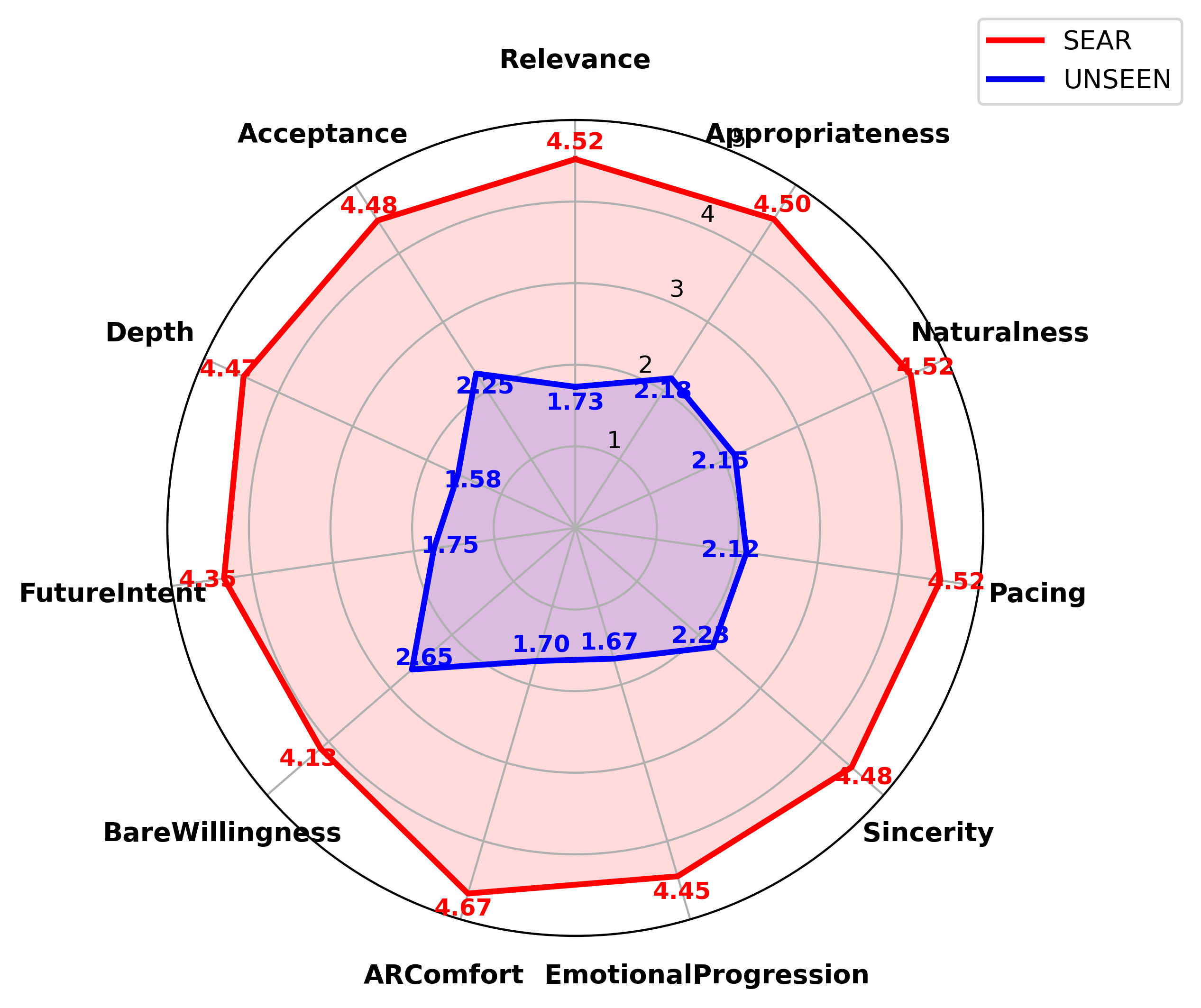}
\caption{Comparison of subjective-experiences.}
\label{fig:experiment_subjective_radar_compare}
\vspace{-10pt}
\end{figure}

\subsection{UNSEEN Subjective Experiences Impact}

Figure~\ref{fig:experiment_subjective_radar_compare} compares SEAR and UNSEEN across 11 subjective-experience dimensions. The trend is consistent: UNSEEN shifts all dimensions from high-risk, high-engagement interaction (SEAR) to low-engagement, low-persuasion interaction, indicating effective disruption of social-engineering dynamics.
Quantitatively, the mean score drops from 4.45 (SEAR average over 11 dimensions) to 1.98 (UNSEEN average), a \textbf{55.5\% reduction}. The largest decreases appear in dimensions that are directly tied to trust manipulation and rapport formation: ARComfort (4.67 $\rightarrow$ 1.70, -63.6\%), Relevance (4.52 $\rightarrow$ 1.73, -61.7\%), Naturalness (4.52 $\rightarrow$ 2.15, -52.4\%), Pacing (4.52 $\rightarrow$ 2.12, -53.1\%), Sincerity (4.48 $\rightarrow$ 2.23, -50.2\%), Depth (4.47 $\rightarrow$ 1.58, -64.7\%), and FutureIntent (4.35 $\rightarrow$ 1.75, -59.8\%).
These reductions show that UNSEEN not only lowers immediate willingness to cooperate, but also weakens the attacker’s ability to sustain believable, emotionally calibrated, multi-turn influence. In other words, the defense suppresses the core mechanisms that make SEAR effective: contextual relevance, conversational fluency, perceived sincerity, and long-term re-engagement.
From a security perspective, this subjective-experience degradation is desirable for defense. SEAR relies on making interactions feel natural, comfortable, and trustworthy to increase compliance; UNSEEN systematically breaks this effect. Combined with the objective attack-channel results in Table~\ref{tab:sear_unseen_score_distribution}, the subjective findings provide converging evidence that UNSEEN effectively defends against AR-LLM-SE attacks.

\section{Conclusion}
\label{sec:conclusion}

This work presents UNSEEN, a coordinated cross-stack defense for AR-LLM-SE risks. UNSEEN combines identity-gated sensing at the AR layer, targeted profile suppression at the model layer, and release-time guardrails at the interaction layer. 
Our evaluation shows that UNSEEN consistently shifts user responses from likely compliance to likely rejection 
against AR-LLM-SE attacks.
%
We hope UNSEEN provides a practical foundation for secure AR-LLM system design, platform policy, and real-world safety evaluation.




%



\newpage

\bibliographystyle{abbrv}
\bibliography{bibs/cgraph}


\end{document}